\documentclass[%
 reprint,
groupedaddress,
amsmath,
amssymb,
aps,
prd,
twocolumn,preprintnumbers,floatfix,nofootinbib]{revtex4-1}

\usepackage{graphicx}% Include figure files
\usepackage{dcolumn}% Align table columns on decimal point
\usepackage{bm}% bold math
\usepackage{hyperref}% add hypertext capabilities
\usepackage{float}
\usepackage{color}

\newcommand{\be}{\begin{equation}}
\newcommand{\ee}{\end{equation}}
\newcommand{\bea}{\begin{eqnarray}}
\newcommand{\eea}{\end{eqnarray}}

\usepackage{mathrsfs}
\usepackage{amsmath, amsthm, amssymb}
\usepackage{color}
\usepackage{epstopdf}
\usepackage{amssymb}
\usepackage{verbatim}
\usepackage{hyperref}
\usepackage{enumerate}
\usepackage{float}
\usepackage[caption = false]{subfig}
\usepackage[normalem]{ulem}  
\usepackage{epsfig}
\usepackage{graphicx}%%%%%%%%%%%
\newcommand{\apjl}{ApJLett}		% Astrophysical Journal, Letters
\newcommand{\jcap}{JCAP}		% JCAP
\newcommand{\physrep}{Phys.~Rep.}   % Physics Reports
\begin{document}

%\preprint{APS/123-QED}
\title{Self-Interaction Controls Vortex Scale in Soliton Mergers}

% more complex case: 4 authors, 3 institutions, 2 footnotes
\author{Yuanyuan Zeng$^1$}
\author{Bokai Zhang$^1$}
\email{zbk329@swu.edu.cn}
\author{and Jiajun Chen$^1$}
\thanks{Corresponding author}
\email{chenjiajun@swu.edu.cn}
\affiliation{
$^1$School of Physical Science and Technology, Southwest University, Chongqing 400715, China
}

% e-mail addresses: one for each author, in the same order as the authors

%Lines break automtically or can be forced with \\

%\affiliation{%


%\affiliation{
%\affiliation{
%\affiliation{%


\date{\today}% It is always \today, today,

\begin{abstract}
 This study investigates the impact of self-interaction strength on the formation and scale of turbulent vortex structures during the merger of Bose stars, using numerical simulations of the Gross-Pitaevskii-Poisson (GPP) equations. We find that vortex formation is a universal outcome of soliton mergers, with the vortex size strongly dependent on the self-interaction coupling parameter $g$. Through analysis of velocity correlations, kinetic energy spectra, and vorticity distributions, we conclude that for repulsive self-interaction, the vortex region expands as self-interaction strength increases; conversely, for attractive self-interaction, the vortex region shrinks as self-interaction strength increases.

\end{abstract}

\maketitle

%\tableofcontents

\section{Introduction}
\label{sec:introduction}
Dark matter (DM) is a theoretical type of matter that comprises roughly 27\% of the universe's total mass-energy \cite{Aghanim:2018eyx}. One widely explored hypothesis suggests that dark matter could be composed of light (pseudo-)scalar particles with large occupation numbers, allowing them to be represented by a classical scalar field $\phi$ \cite{Dine:1982ah,Suarez:2013iw,Preskill:1982cy,Abbott:1982af,2015PhRvD..92j3513G,Widrow:1993qq,Uhlemann:2014npa}. For small values of the field, the potential of the scalar field can be expressed as \cite{Sikivie:2006ni,Arvanitaki:2009fg,2017PhRvL.118a1301L}
\begin{equation}
V(\phi) = \frac{1}{2} m^2 \phi^2 \pm \frac{1}{4!} \frac{m^2}{f_a^2} \phi^4 + \cdots,
\end{equation}
where $m$ is the mass of the particle, $f_a$ represents the decay constant, and natural units ($\hbar = c = 1$) are adopted. The $\pm$ sign in front of the $\phi^4$ term indicates whether the self-interaction is repulsive or attractive. In this study, we introduce the dimensionless self-coupling constant as $g \equiv \pm \frac{1}{8 f_a^2}$. The values of the particle mass and coupling constant differ depending on the specific model. For example, QCD axions are characterized by masses ranging from $10^{-11}$ to $10^{-2}$ eV \cite{pecceiquinn1977,weinberg1978,wilczek1978,Kim:1979if,Shifman:1979if,Dine:1982ah,Zhitnitsky:1980tq,1981PhLB..104..199D,2015PhRvD..91h4011A,2015JCAP...02..006P,Marsh:2015xka,Tanabashi:2018oca,luzio2020landscape}, while ultralight bosons have masses in the range of $10^{-22}$ to $10^{-19}$ eV \cite{1990PhRvL..64.1084P,2000PhRvD..62j3517S,hu2000,2000ApJ...534L.127P}. Previous studies indicate that, for bosons without self-interaction, boson stars undergo mergers under gravitational influence, and coherence effects lead to turbulence formation. The coherence effects of waves lead to turbulence formation, as the fluid velocity results in an irrotational flow, except in cases where the field $\psi$ is discontinuous or lacks first or second derivatives\cite{Mocz:2017wlg}.
Vortex lines can act as gravitational seeds that induce the condensation of ordinary matter, forming persistent ring-like structures and providing an observable probe for detecting this model\cite{Alvarez-Rios:2025ydz}. 
In strong gravitational lensing systems, the interference substructures induced by dark matter vortices lead to 5--10\% flux anomalies, thereby allowing the detection of vortex features, such as the universal profiles of density $\rho \sim r^2$ and velocity $v \sim 1/r$, through observations of their magnification and deflection effects on background sources\cite{Hui_2021}.
On the other hand, self-interactions, whether attractive or repulsive, can substantially influence the dynamics of bosonic systems. Attractive interactions may lead to more compact boson stars, while repulsive ones could increase their radius\cite{PhysRevD.104.083022,PhysRevD.106.023009}. Therefore, it is essential to investigate the impact of self-interaction on the structure of turbulence.

Using pseudo-spectral methods~\cite{Fornberg1987,PhysRevD.106.023009,PhysRevD.108.083021,Chen_2024,PhysRevD.111.043031,PhysRevD.104.083022,Mocz:2017wlg}, we study the self-interaction effects on vortex scaling in soliton mergers. This investigation employs numerical simulations of the Gross-Pitaevskii-Poisson (GPP) equations to explore the formation and scale of turbulent vortex structures during the merger of Boson stars. Our analysis reveals that vortex formation is a universal outcome of soliton mergers, with the vortex size strongly dependent on the self-interaction coupling parameter $g$.

For velocity correlations, we observe that the flow exhibits a pronounced dependence on $g$. At low or negative $g$ values, the velocity field shows short-range correlations with rapid decorrelation. As $g$ increases to positive values, the correlation length extends significantly, indicating enhanced spatial coherence and the emergence of large-scale organized motion patterns.
The kinetic energy spectrum further supports this trend. For negative $g$, the spectral peak occurs at high wavenumbers with a high energy amplitude, reflecting small-scale fluctuations. With increasing $g$ to positive values, the peak shifts to lower wavenumbers and the amplitude decreases, indicating a suppression of small-scale energy and a dominance of large-scale coherent structures.
Vorticity distributions show that the spatially averaged vorticity $( \langle |\omega| \rangle = \langle| \nabla \times\bm{v} |\rangle )$ decreases with increasing $g$, with a subtle non-monotonic bump at small positive $g$ due to transient increases in local shear. The probability distribution function (PDF) of vorticity magnitude remains Gaussian across all $g$, suggesting statistical homogeneity despite coherent structures. Vortex size distributions narrow with increasing $g$, reflecting a well-defined characteristic size and a reduction in small-scale variability, consistent with the growth of large-scale coherence driven by nonlinear self-interactions.

We begin in Sec.~\ref{sec:equations} by introducing the GPP equations and our numerical setup with varying soliton configurations. In Section~\ref{sec:merger}, we investigate the formation of turbulent structures in the process of the merger of solitons by analyzing velocity correlations and the kinetic energy spectrum under varying self-interaction strengths, as well as by examining vorticity distributions and vortex size scaling. Finally, we present our conclusions in Sec.~\ref{sec:conclusion}.

\section{Field equations and numerical setup}
\label{sec:equations}
In the non-relativistic, low-density and low-velocity limits, we can rewrite the scalar field $\phi$ as
\begin{equation}
\phi= \sqrt{\frac{2}{m}} {\rm Re}(\psi e^{-i m t}).
\label{eq:phi_psi}
\end{equation}
The complex field $\psi$ at lowest order satisfies the GPP equations
\cite{Chavanis:2011zm,Eby:2015hsq,Salehian:2021khb}
\begin{eqnarray}
i\frac{\partial}{\partial{t}}\psi&=&-\frac{1}{2m}\nabla^2\psi + m V\psi+g|\psi|^2\psi,
\label{eq:GPP1}
\\
\nabla^2{V}&=&4 \pi G m\left(|\psi|^2-n\right),
\label{eq:GPP2}
\end{eqnarray}
where where $\psi$ is a nonrelativistic classical field, $G$ is Newton's gravitational constant, $V$ is the gravitational potential, $g$ is the dimensional self-interaction coupling, and $n$ is the mean number density.

The mass density $\rho$ and a velocity field $\mathbf{v}$ can be obtained by introducing the Madelung transformation to the field, $\psi = \sqrt{\rho}e^{iS}$, where the velocity is given by $\mathbf{v}(\mathbf{r},t) = \nabla S(\mathbf{r},t)/m$, $S$ is the phase.
One thus obtains \cite{PhysRevD.84.043531} a continuity equation for $\rho$ in the form
\begin{equation}
\frac{\partial}{\partial t}\rho + \nabla\cdot(\rho\mathbf{v}) = 0
\end{equation}
and a corresponding equation for the velocity $\mathbf{v}$
\begin{equation}
\frac{\partial}{\partial t}\mathbf{v} + \frac{\nabla}{m}\left[\frac{m|\mathbf{v}|^2}{2} + g\rho + V - \frac{\hbar^2}{2m}\frac{\nabla^2\sqrt{\rho}}{\sqrt{\rho}}\right] = 0,
\end{equation}
expressed in terms of self-interaction, gravitational potential and quantum pressure contributions. Eqs. (\ref{eq:GPP1}) and (\ref{eq:GPP2}) can be written in a dimensionless form following the definitions in Ref.~\cite{Levkov:2018kau}: substitutions:
\begin{align}
\label{nondimensionalize}
    &x \rightarrow \widetilde{x}/(m v_0), \quad 
    t \rightarrow \widetilde{t}/(m v_0^2), \quad 
    V \rightarrow \widetilde{V} v_0^2, \notag \\
    &\psi \rightarrow \widetilde{\psi} v_0^2 \sqrt{m/(4 \pi G)}, \quad 
    g \rightarrow \widetilde{g} 4\pi G/v_0^2,
\end{align}
where $v_0$ is a reference velocity, e.g. the characteristic velocity of the initial state.
The dimensionless equations are given by
\begin{eqnarray}
i\frac{\partial}{\partial{\widetilde{t}}}\widetilde{\psi}&=&-\frac{1}{2}\widetilde{\nabla}^2\widetilde{\psi} + \widetilde{V}\widetilde{\psi}+\widetilde{g}|\widetilde{\psi}|^2\widetilde{\psi},
\label{eq:GPP1_dim}
\\
\widetilde{\nabla}^2{\widetilde{V}}&=&|\widetilde{\psi}|^2-\widetilde{n},
\label{eq:GPP2_dim}
\end{eqnarray}

To simulate the GPP equations Eq.\ref{eq:GPP1} and Eq.\ref{eq:GPP2}, we employ a fourth-order pseudospectral method as described in \cite{Du:2018qor,PhysRevD.104.083022}. For the initial condition, we consider initial conditions with varying soliton configurations: the number of solitons ranges from 4 to 50, with different soliton mass and initial positions.

\section{Statistical and Structural Properties of Turbulence}
\label{sec:merger}

\begin{figure}[htbp]
\centering

\includegraphics[width=\columnwidth]{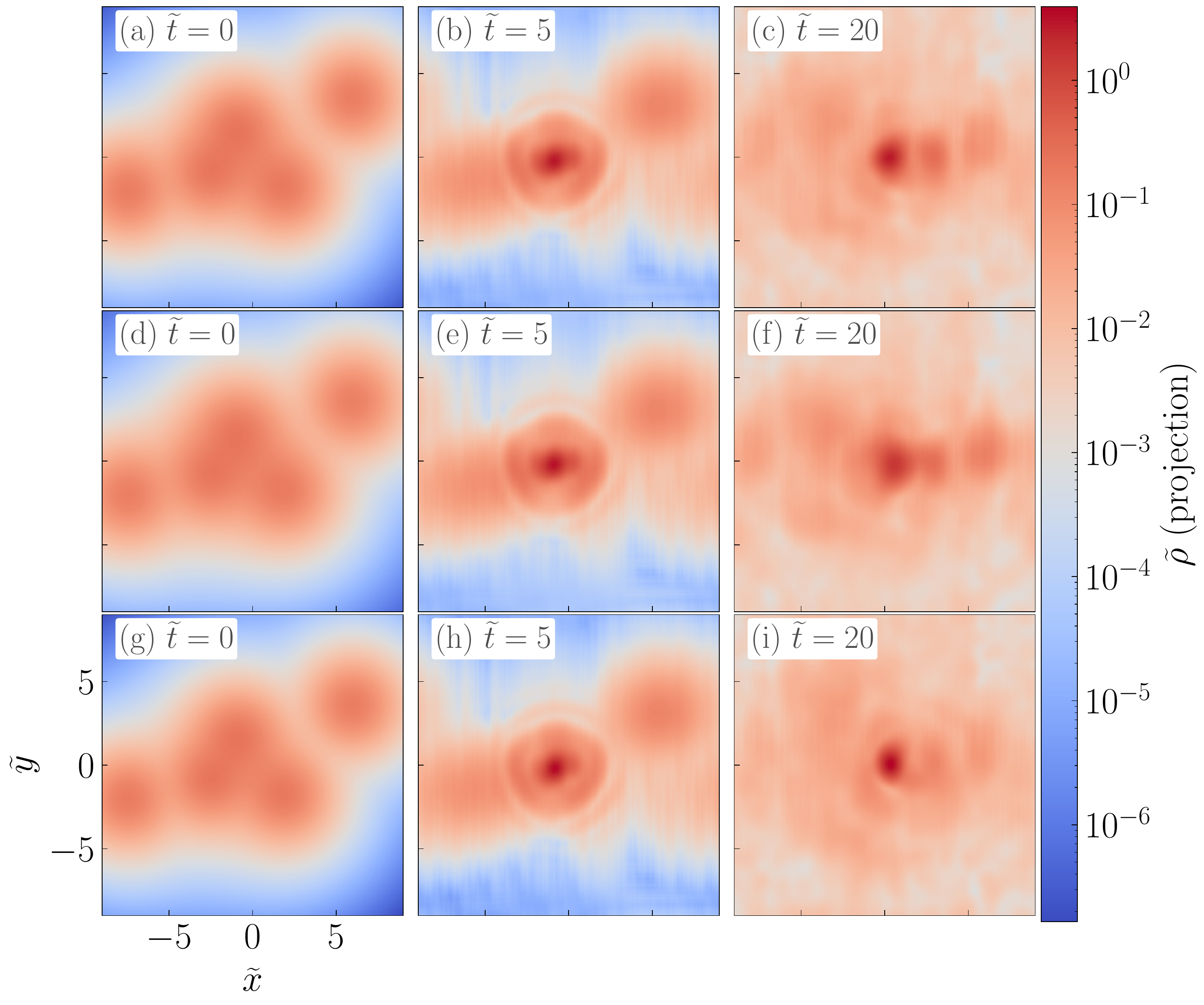}
\caption{Snapshots of the density field at different time from simulations in a box of size $\widetilde{L}=18$ and total mass $\widetilde{N} = 157$. Upper row is $\widetilde{g} = 0$. Middle row is $\widetilde{g} = 0.01$. Lower row is $\widetilde{g} = -0.01$.}
\label{fig:rho_projection}
\end{figure}

\begin{figure}[htbp]
\centering
\includegraphics[width=\linewidth]{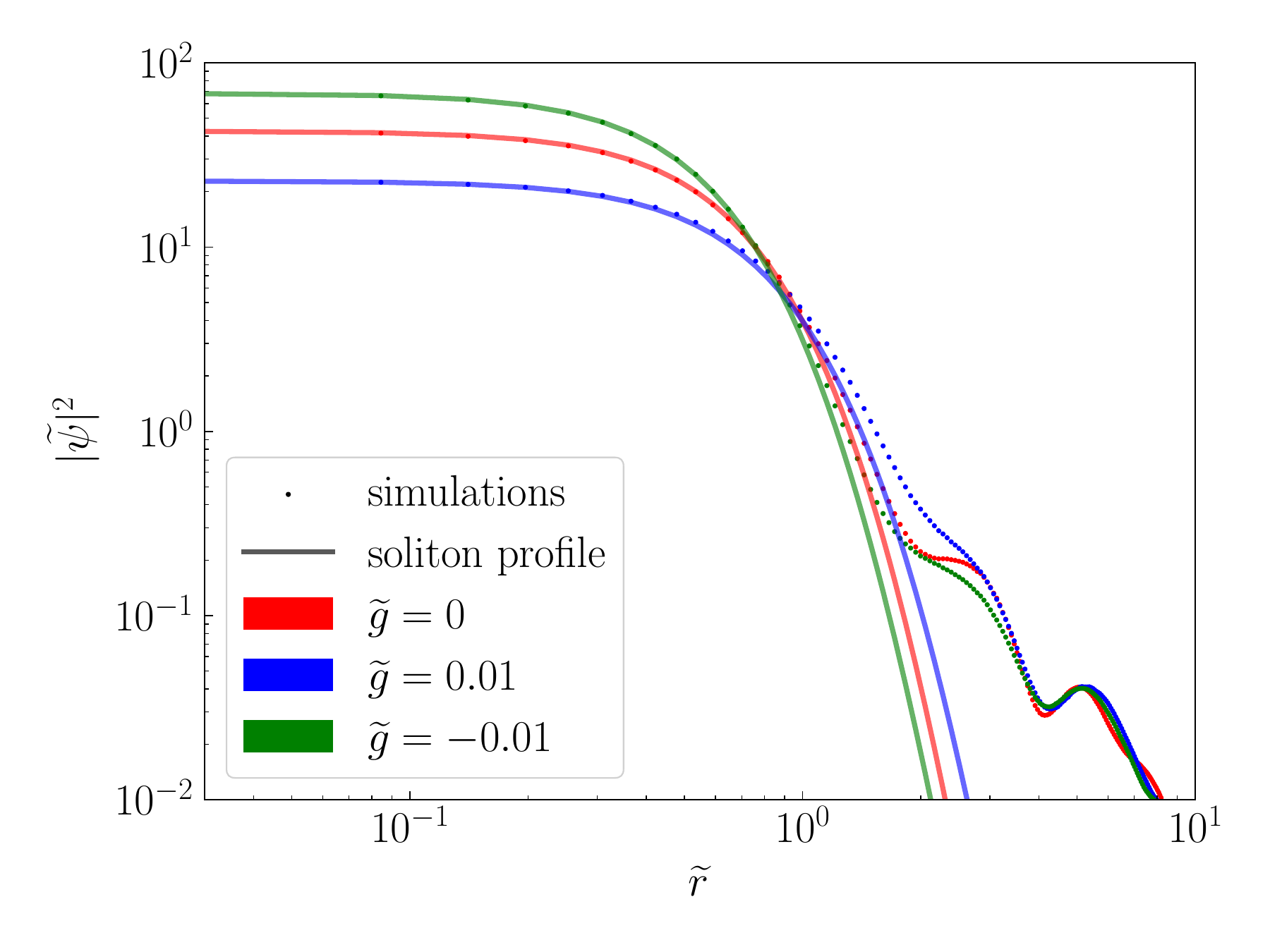}
\caption{Density profiles of the halos from simulations (colored dots) with $\widetilde{g} = 0$, $\widetilde{g} = 0.01$, $\widetilde{g} = -0.01$, compared with solitonic profiles (solid lines) as given by \cite{Schive:2014hza,PhysRevD.104.083022}}
\label{fig:rho_profile}
\end{figure}

We run a large number of simulations of soliton mergers using initial conditions generated for systems containing 4 to 50 solitons in a box, based on the fitting function for solitons with or without self-interaction \cite{Schive:2014hza,PhysRevD.104.083022}. All simulations were run with identical soliton masses but varying values of $\widetilde{g}$. Several simulations is illustrated in Fig.~\ref{fig:rho_projection}, where we observe the solitons gradually merging and eventually forming a single soliton surrounded by a halo\cite{PhysRevD.104.083022,Zhang:2024bjo,Mocz:2017wlg,Schive:2014dra}. The radial density profiles of the halo, along with the analytic profiles of solitons with and without self-interactions, are presented in Fig.~\ref{fig:rho_profile}. These profiles are well-fitted by the soliton solution)\cite{Schive:2014hza}, matching the central number density (solid lines). Fig.~\ref{fig:rho_projection} and Fig.~\ref{fig:rho_profile} show that their density distributions are not exactly the same, so it is necessary for us to analyze the distribution state of turbulence under different self-interactions.

\subsection{Velocity correlations}
\label{sec:velocity}

The hallmark of classical turbulence is the spontaneous emergence of intricate vortex structures and highly unpredictable, spatiotemporally chaotic flows, quantitatively characterized by the local vorticity field $\boldsymbol{\omega} = \nabla \times \mathbf{v}$. These complex, multiscale structures typically arise in high Reynolds number regimes and serve as the dominant mechanism for the highly efficient transport of momentum and energy across a broad spectrum of scales in classical fluids. In stark contrast, quantum fluids exhibit fundamentally different underlying dynamics rooted in the macroscopic wave nature of their constituent order parameter. Specifically, the superfluid velocity is given by the gradient of the phase of the macroscopic field, $\mathbf{v} = \nabla S / m$. This fundamental definition intrinsically implies that the flow is rigorously irrotational everywhere except at isolated singularities---discrete points or continuous lines where the density precisely vanishes or its spatial derivatives become undefined. Consequently, vorticity in quantum fluids is not continuously distributed throughout the bulk, but is instead quantized and rigorously confined to ultrathin vortex cores: topological defects around which the phase of the field winds by integer multiples of $2\pi$ and the fluid density asymptotically approaches zero. Although the velocity formally diverges in the immediate vicinity of these vanishingly small cores, the local vanishing density ensures that all physically observable quantities, such as kinetic energy density, remain perfectly finite and well-behaved.

In this comprehensive study, we conduct an in-depth investigation into the decisive influence of nonlinear self-interactions, characterized by the coupling parameter $g$, on the three-dimensional velocity field structure of the system. Across the explored range of $g$ values, our simulations reveal that while the overall presence of vortical motion is a robust feature, the spatial organization and collective nature of the flow exhibit a pronounced dependence on the self-interaction strength. As $g$ increases, the velocity field progressively develops stronger spatial coherence, with localized fluctuations giving way to large regions of correlated motion. This trend highlights the capacity of nonlinear interactions to restructure the flow from a state dominated by small, irregular eddies into one exhibiting extended, organized motion patterns.

\begin{figure*}
\centering
\includegraphics[width=2.0\columnwidth]{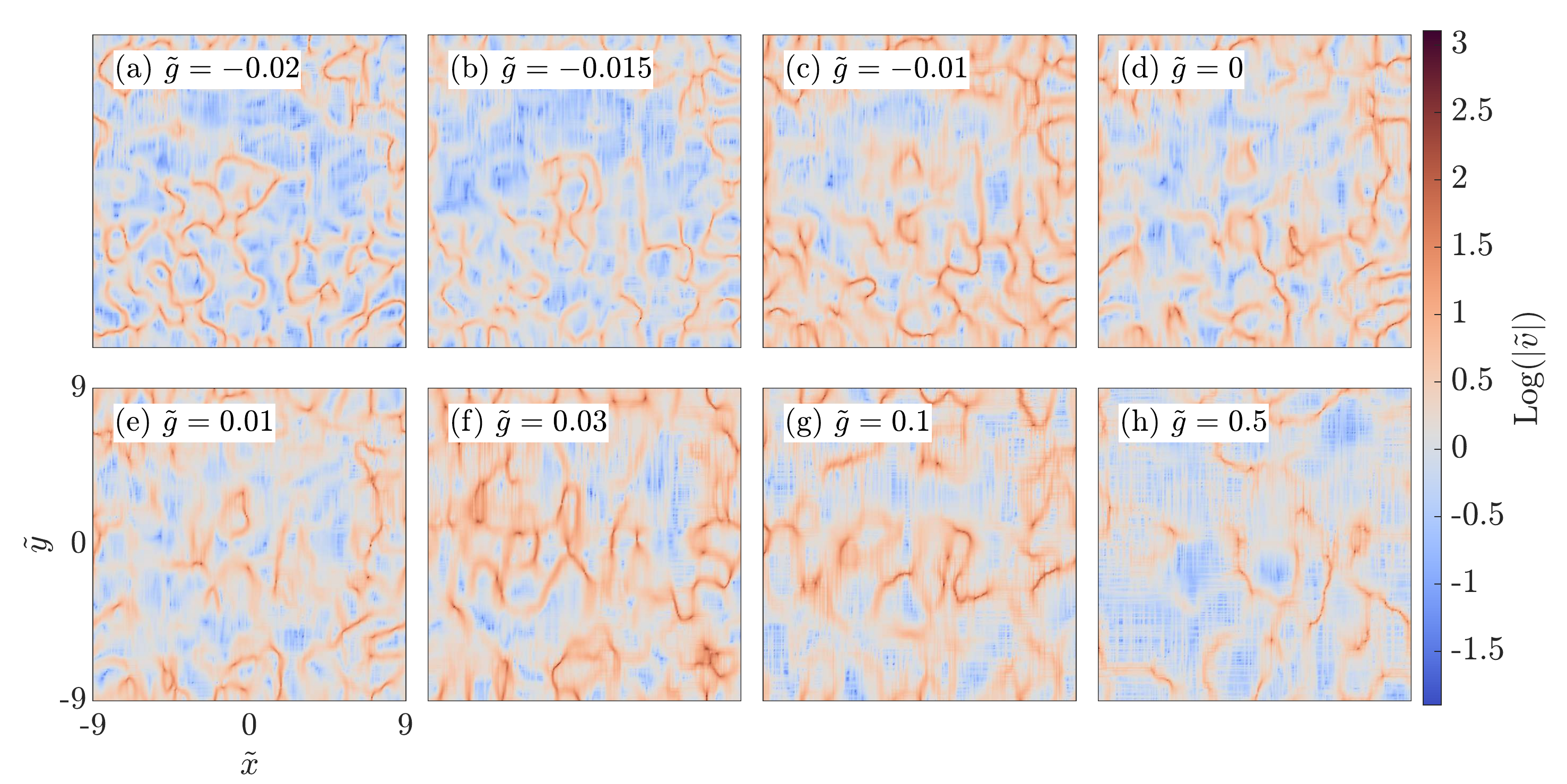}
\caption{Velocity field snapshots at various self-interaction coupling strengths $\widetilde{g}$. The color map represents the magnitude of the velocity $|\bm{v}|$.}
\label{fig:VelocityField}
\end{figure*}

Detailed snapshots of the three-dimensional velocity field, as shown in Fig.~\ref{fig:VelocityField}, clearly illustrate this transformation.

At low $\widetilde{g}$ values (e.g., $\widetilde{g} = -0.02$), the flow appears highly disordered, with velocity vectors rapidly changing direction over short distances, indicative of weak long-range correlations and a prevalence of small-scale, uncoordinated structures. As g is increased to positive values (e.g., $\widetilde{g} = +0.5$), the flow undergoes a striking reorganization: velocity vectors become increasingly aligned over larger spatial extents, and coherent motion dominates a significant fraction of the domain. This evolution is accompanied by a systematic growth in the mean flow amplitude, reflecting a net increase in kinetic energy driven directly by the nonlinear self-interaction term.

This marked enhancement of collective velocity correlations at higher g suggests that nonlinear self-interaction acts as a powerful mechanism for promoting large-scale organization in the flow. The emergence of coherent, domain-spanning velocity structures indicates an efficient transfer of kinetic energy from small to large spatial scales, reminiscent of an inverse energy cascade in three-dimensional settings with suppressed vortex stretching. Consequently, the self-interaction term can be regarded as an active driver of macroscopic flow order, fostering persistent, spatially extended circulation patterns and significantly amplifying both the magnitude and the spatial range of velocity correlations in the system.

The elongated structures visible in the three-dimensional velocity field suggest an interesting analogy with the transverse profiles of stationary spinning Q-balls ~\cite{Volkov:2002aj,Galushkina:2025rotating}. Although the underlying equations are different, both systems exhibit coherent nonlinear self-organized structures, and this resemblance may be useful for interpreting the geometry of the flow configurations observed here.

\begin{figure}
\centering
\includegraphics[width=\linewidth]{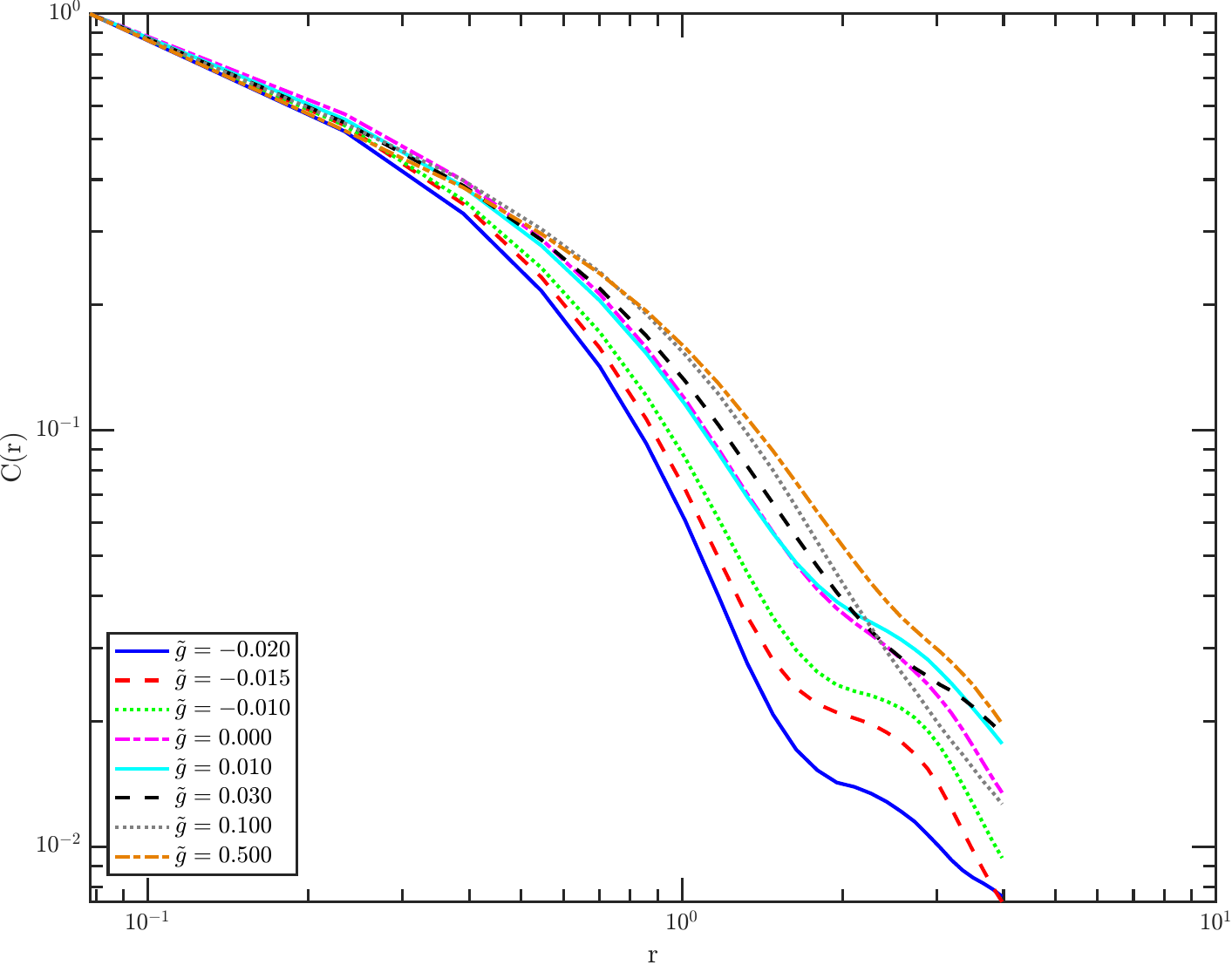}
\caption{The log-log plot of velocity correlation functions at different strengths $\widetilde{g}$.}
\label{fig:velocity_correlation}
\end{figure}

A complementary and highly informative perspective on the flow organization is provided by the velocity correlation function (VCF), which quantifies the degree of spatial coherence in the velocity field. The VCF, shown graphically in Fig.~\ref{fig:velocity_correlation}, is defined as \cite{Yeomans2013}

\begin{equation}
C_v(r) = \frac{\langle \mathbf{v}(\mathbf{x}) \cdot \mathbf{v}(\mathbf{x}+\mathbf{r}) \rangle}{\langle |\mathbf{v}(\mathbf{x})|^2 \rangle},
\end{equation}

where $\mathbf{v}(\mathbf{x})$ denotes the velocity vector at position $\mathbf{x}$, $\mathbf{r}$ is the separation vector, and the angle brackets represent an average over all spatial positions and, if applicable, over time. By construction, $C_v(0) = 1$, and the function systematically decays as $r$ increases, capturing the gradual loss of local velocity correlation over larger distances. Both the separation $r$ and the correlation $C_v(r)$ are often plotted on logarithmic scales to better visualize the decay over multiple spatial decades.

For three-dimensional velocity fields, $C_v(r)$ is computed by first calculating the dot product between velocity vectors separated by a distance $r$ along all directions in the domain, and then averaging over all pairs of points separated by that distance. Practically, this is implemented by binning the separation distances and computing  

\begin{equation}
C_v(r_i) = \frac{\sum_{\mathbf{x}, \mathbf{r} \in \text{bin } i} \mathbf{v}(\mathbf{x}) \cdot \mathbf{v}(\mathbf{x}+\mathbf{r})}{\sum_{\mathbf{x}} |\mathbf{v}(\mathbf{x})|^2}.
\end{equation}

This approach ensures that the correlation function captures the isotropic properties of the flow while providing robust statistical averaging over the entire three-dimensional domain.

The decay behavior of the VCF exhibits a pronounced dependence on the self-interaction parameter g. For weak or negative interactions, such as $\widetilde{g} = -0.020$, $C_v(r)$ drops rapidly, reaching a value of approximately 0.1 at a relatively short separation $r \approx 0.9$, indicating that the flow is primarily dominated by short-range correlations and small-scale velocity fluctuations. In contrast, for stronger positive interactions, such as $\widetilde{g} = +0.500$, the same level of decorrelation occurs only at a substantially larger separation $\widetilde{r} \approx 3.0$, reflecting a significantly extended correlation length and the emergence of coherent, long-range flow structures.  

This monotonic enhancement of spatial coherence with increasing g clearly demonstrates that stronger self-interactions promote collective organization in the velocity field. The VCF not only provides a quantitative measure of the correlation length but also reveals how nonlinearity facilitates the persistence of long-range structures, enabling macroscopic flows to develop robust patterns that span a substantial portion of the system. Detailed analysis of the VCF decay profile can be further used to extract characteristic length scales and identify crossover points between short-range fluctuation-dominated regions and long-range coherent domains, providing a rigorous framework for connecting microscopic interactions to emergent large-scale flow organization.

\subsection{Kinetic Energy Spectrum}
\label{sec:spectrum}

The kinetic energy spectrum, $E(k)$, is computed from the three-dimensional velocity field $\mathbf{v}(\mathbf{x})$ obtained in the simulations. The calculation follows the standard procedure \cite{Giomi2015}:

\begin{equation}
E(k) = \frac{1}{2} \sum_{k \leq |\mathbf{k}'| < k + \Delta k} |\hat{\mathbf{v}}(\mathbf{k}')|^2,
\end{equation}

where $\hat{\mathbf{v}}(\mathbf{k}')$ denotes the Fourier transform of the velocity field:

\begin{equation}
\hat{\mathbf{v}}(\mathbf{k}) = \frac{1}{V} \int \mathbf{v}(\mathbf{x}) e^{-i \mathbf{k} \cdot \mathbf{x}} d^3x,
\end{equation}

with $V$ representing the volume of the simulation domain. The sum runs over all Fourier modes whose magnitude lies within the interval $[k, k + \Delta k]$, where $\Delta k$ defines the width of the wavenumber bin. In practice, the three-dimensional velocity field is transformed via FFT, the squared modulus of each Fourier mode is computed, and the resulting values are binned according to the magnitude of the corresponding wavevector to construct an isotropic spectrum. Time averaging over multiple snapshots is performed to minimize statistical fluctuations.

\begin{figure}
\centering
\includegraphics[width=\linewidth]{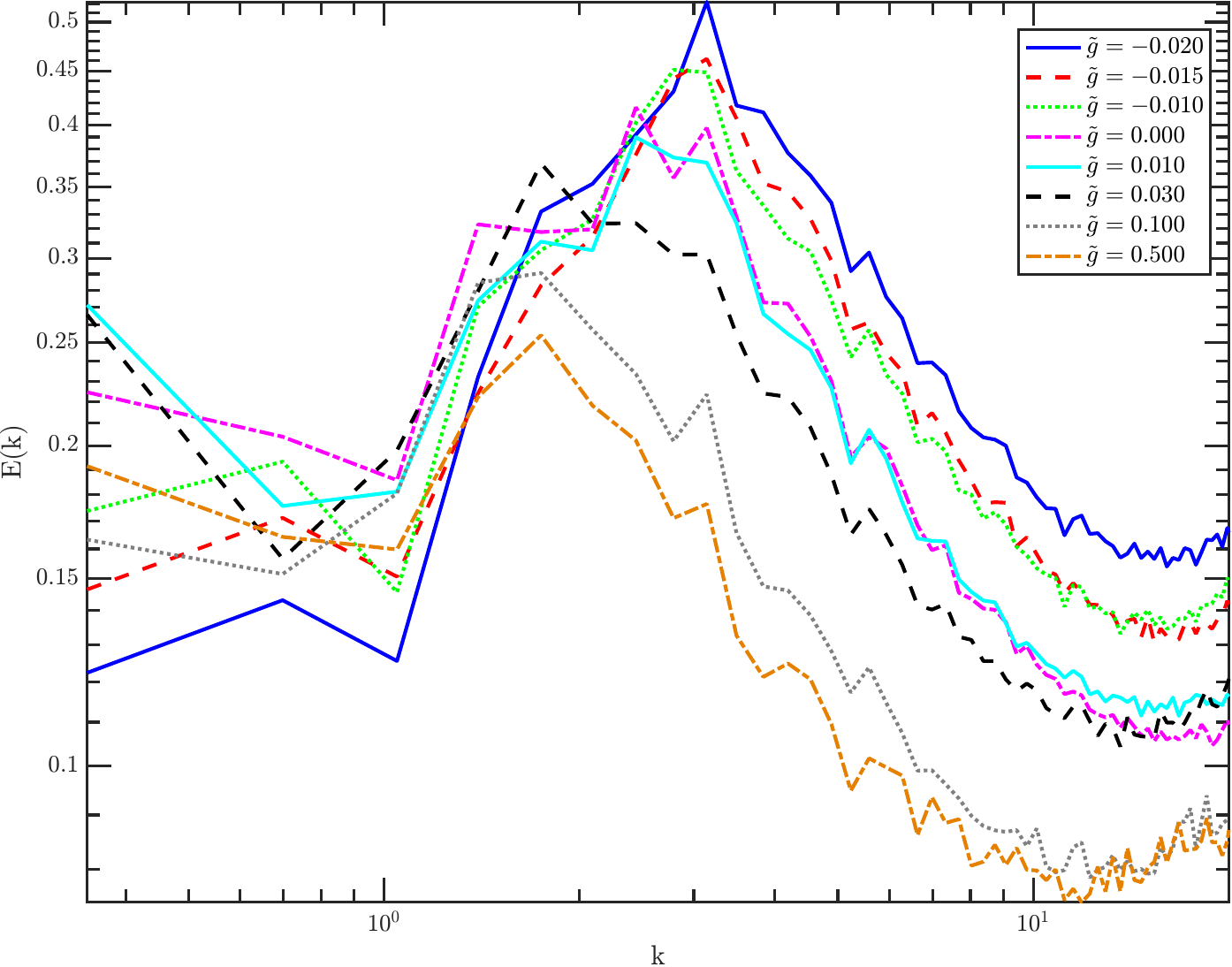}
\caption{Normalized energy spectrum at $\widetilde{t}=20$ with different $\widetilde{g}$.}
\label{fig:energy_spectrum}
\end{figure}

The kinetic energy spectra, presented in Fig.~\ref{fig:energy_spectrum}, provide a rigorous quantitative assessment of how the parameter $\widetilde{g}$ modulates the distribution of kinetic energy across spatial scales. 
Each spectrum exhibits a distinct peak, representing the wavenumber at which energy injection, nonlinear transfer, and dissipation approximately reach equilibrium. Wavenumbers below the peak correspond to energy-containing scales where nonlinear interactions can facilitate the transfer of energy to larger structures, whereas wavenumbers above the peak correspond to scales dominated by dissipation. The spatial scale associated with the spectral peak, approximately given by $l_{\rm peak} \sim 1/k_{\rm peak}$, aligns closely with the velocity correlation length derived from the velocity correlation function, providing a direct connection between spectral properties and real-space flow coherence.

For negative values of $\widetilde{g}$  (e.g., $\widetilde{g} = -0.020, -0.015, -0.010$), the spectral peaks are located at relatively high wavenumbers ($k_{\rm peak} \approx 3$), indicating that kinetic energy is concentrated at small scales and that the velocity field exhibits short-range spatial coherence. In particular, the case $\widetilde{g} = -0.020$ displays the largest peak, $E(k_{\rm peak}) \approx 0.5$, highlighting the pronounced presence of fine-scale, energetic fluctuations. As $\widetilde{g}$ increases to positive values, the spectral peak progressively shifts toward lower wavenumbers ($k_{\rm peak} \approx 2$ for $\widetilde{g} = +0.500$), corresponding to larger characteristic spatial scales. Concurrently, the overall amplitude of the spectrum across most wavenumbers decreases, reflecting the attenuation of small-scale fluctuations and the concomitant emergence of large-scale coherent structures.

The observed correspondence between $k_{\rm peak}$ and the velocity correlation function underscores that the spatial coherence of the velocity field grows systematically with increasing $g$. Larger values of $g$ favor extended correlation lengths and the persistence of coherent, long-lived flow structures. These findings are consistent with the qualitative features observed in three-dimensional snapshots of the velocity field, which reveal that increasing $g$ leads to increasingly collective motions, well-defined vortical structures, and enhanced long-range correlations. Collectively, the kinetic energy spectra and velocity correlations demonstrate that nonlinear self-interactions play a decisive role in organizing the flow: they suppress small-scale fluctuations while simultaneously promoting the formation and maintenance of large-scale coherent motions, establishing a clear link between microscopic interaction strength and emergent macroscopic flow organization.

\subsection{Vorticity Structures}
\label{sec:vorticity}

In our simulations, the three-dimensional velocity field $\bm{v}(\bm{x})$ is obtained on a uniform grid. The vorticity $\bm{\omega}(\bm{x})$ is computed numerically using a central finite-difference scheme to evaluate the spatial derivatives along each coordinate direction. The resulting vorticity field provides a detailed map of the local rotational motion within the fluid. For statistical analysis, both the magnitude of the vorticity $|\bm{\omega}|$ and its components are considered.  

\begin{figure}
\centering
\includegraphics[width=\linewidth]{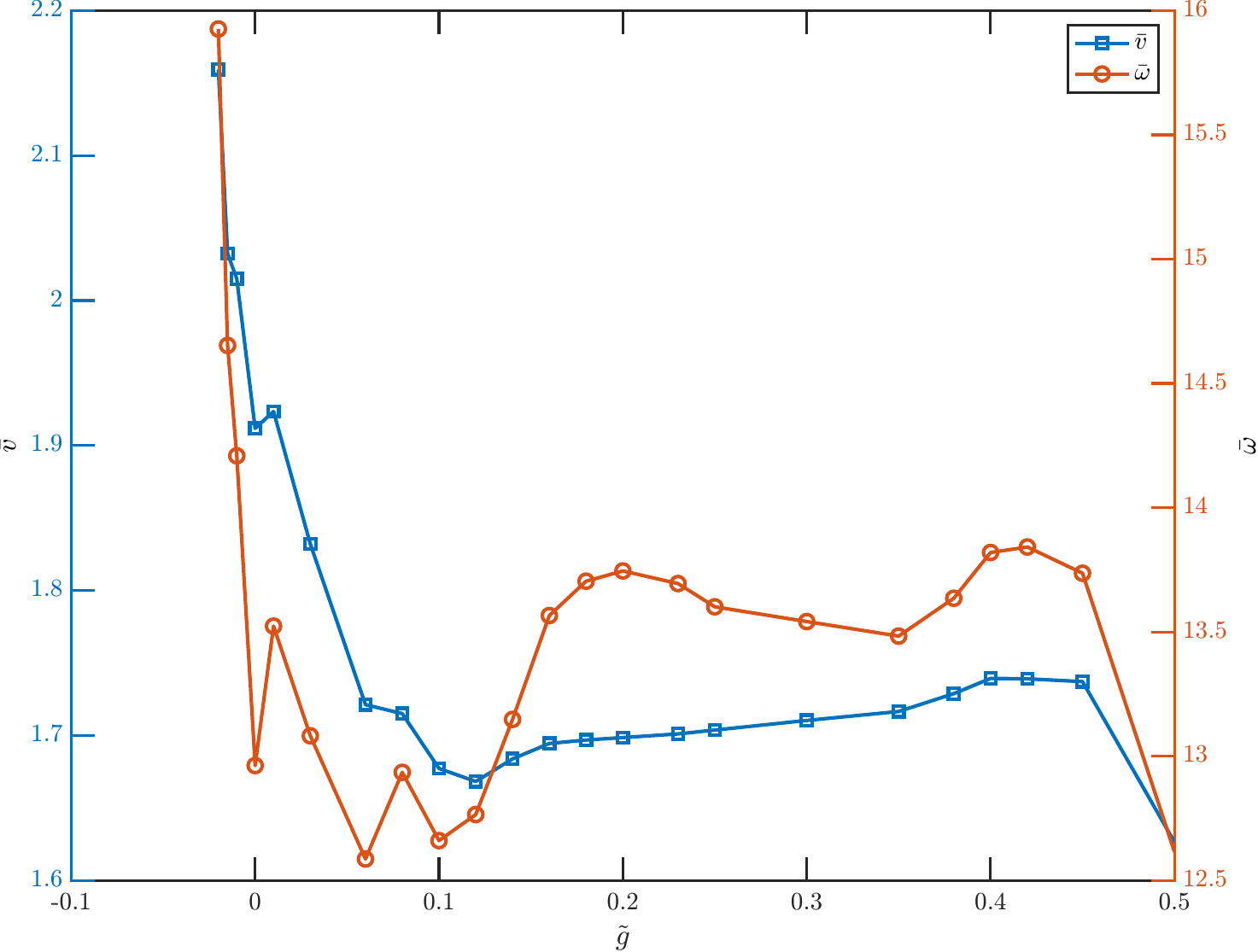}
\caption{Mean vorticity (right axis) and mean velocity (left axis) as a function of self-interaction strength.}
\label{fig:VorticityAve}
\end{figure}

The spatially averaged vorticity, $\langle |\bm{\omega}| \rangle$, exhibits a nontrivial dependence on the self-interaction parameter g, as shown in Fig.~\ref{fig:VorticityAve}. Overall, $\langle |\bm{\omega}| \rangle$ decreases with increasing g, indicating that stronger self-interactions suppress local rotational motions. However, this decrease is not strictly monotonic. In the intermediate range of g, the mean vorticity shows a noticeable variation, including a mild increase, before decreasing again at larger g. This behavior suggests a competition between the formation of coherent large-scale flow structures and the suppression of fine-scale turbulent fluctuations: at moderate values of g, the increasing flow organization can enhance local shear and vorticity, whereas at larger g the suppression of small-scale vortical activity becomes dominant, leading to an overall reduction of $\langle |\bm{\omega}| \rangle$.

Comparing the trend of decreasing mean vorticity with the earlier observation of increasing velocity correlation lengths reveals a clear physical insight: as g increases, the flow becomes more spatially organized and collective, so the fluid motion is dominated by coherent structures rather than random small-scale eddies. Although the vorticity does not decrease monotonically over the entire parameter range, its overall reduction at larger g accompanies the growth of long-range flow correlations. This indicates a progressive redistribution of kinetic activity from small-scale rotational motions toward larger-scale coherent circulation, consistent with the emergence of increasingly organized flow patterns in the system.

\begin{figure}
\centering
\includegraphics[width=\linewidth]{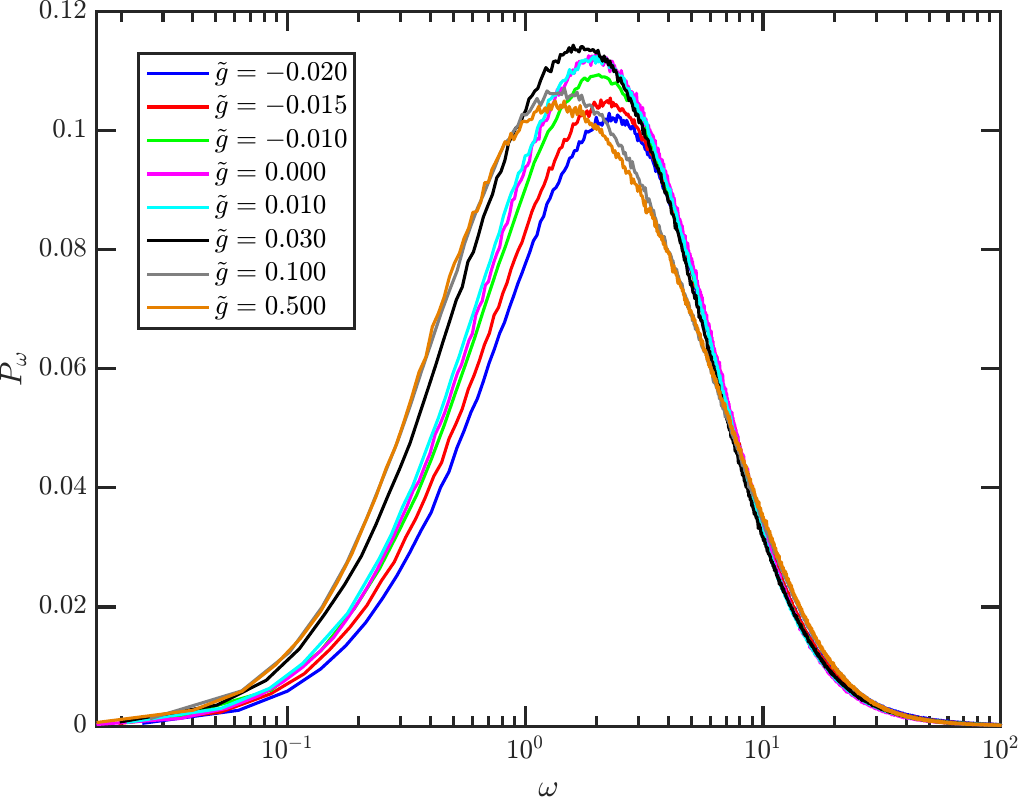}
\caption{Probability distribution functions of vorticity at different self-interaction strengths.}
\label{fig:VorticityPDF}
\end{figure}

The probability distribution function (PDF) of the vorticity magnitude, shown in Fig.~\ref{fig:VorticityPDF}, closely follows a Gaussian profile for all values of g. This observation indicates that, despite the presence of coherent structures at larger g, local vorticity fluctuations remain approximately normally distributed, reflecting the statistical homogeneity of the turbulent field at small scales.

To further characterize coherent rotational structures, we employ the three-dimensional $\lambda_2$- or Okubo--Weiss (OW) method to identify vortical regions and compute their sizes. In this work, we specifically use the $Q$-criterion, which quantifies the local balance between rotation and strain in the flow, and is defined as  
\begin{equation}
Q = \frac{1}{2} \left( \| \boldsymbol{\Omega} \|^2 - \| \mathbf{S} \|^2 \right),
\end{equation}
where $\mathbf{S}$ is the symmetric strain-rate tensor,
\begin{equation}
S_{ij} = \frac{1}{2} \left( \frac{\partial v_i}{\partial x_j} + \frac{\partial v_j}{\partial x_i} \right),
\end{equation}
which measures the rate of deformation of fluid elements without rotation, while $\boldsymbol{\Omega}$ is the antisymmetric rotation-rate tensor,
\begin{equation}
\Omega_{ij} = \frac{1}{2} \left( \frac{\partial v_i}{\partial x_j} - \frac{\partial v_j}{\partial x_i} \right),
\end{equation}
which represents the local rigid-body rotation of the fluid. The norms $\|\mathbf{S}\|$ and $\|\boldsymbol{\Omega}\|$ are computed as the Frobenius norms of the corresponding tensors.

A grid point is identified as belonging to a vortex core if $Q>0$, meaning that the local rotational motion dominates over strain deformation. To separate individual vortex structures, we apply a three-dimensional 26-neighbor connectivity analysis, where two points are considered connected if they share a face, an edge, or a corner. This procedure ensures that all voxels forming a continuous rotational structure are grouped into the same object, even if the connection is only diagonal in the grid.

Once all connected vortical structures are identified, we compute their sizes by counting the number of grid points $N$ in each structure. The physical volume of a vortex is then obtained as
\begin{equation}
v_w = N \times \Delta V,
\end{equation}
where $\Delta V$ is the volume of a single grid cell. This enables a direct statistical analysis of the vortex size distribution across different flow conditions.

\begin{figure}
\centering
\includegraphics[width=\linewidth]{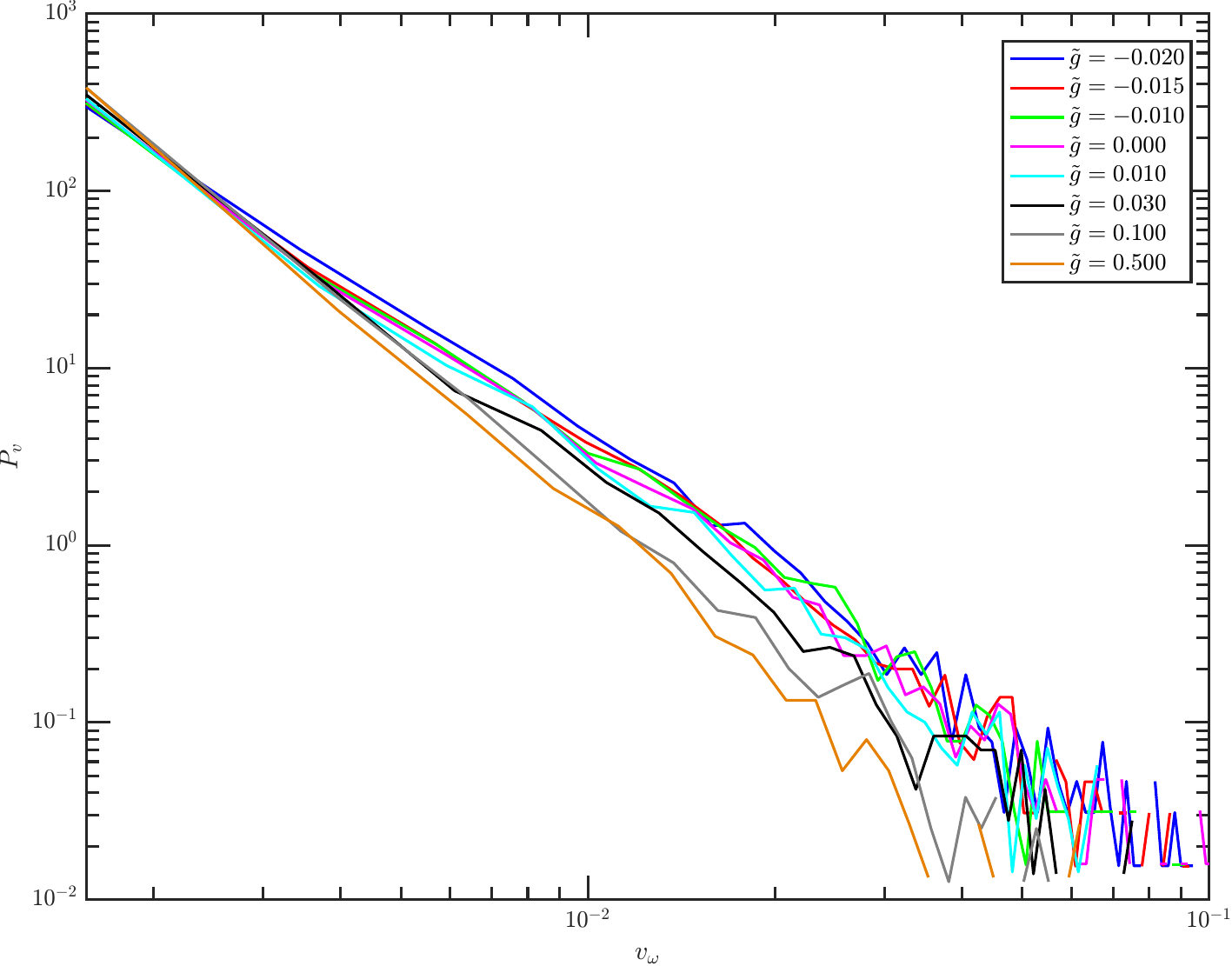}
\caption{Probability distribution function of 3D vortex volume for different self-energy interactions.}
\label{fig:vortexSize}
\end{figure}

The vortex size distribution, shown in Fig.~\ref{fig:vortexSize}, exhibits a pronounced dependence on the interaction strength $\widetilde{g}$. 
As $\widetilde{g}$  increases, the distribution becomes progressively narrower, indicating the emergence of a well-defined characteristic vortex size within the flow. In contrast, at lower $\widetilde{g}$ the distribution is broader, reflecting a wider range of spatial scales. The tail of the distribution deviates slightly from a pure Gaussian form, showing a mild heavy-tailed behavior that can be approximated by a generalized Gaussian with an exponent slightly below $2$. This tail behavior implies the presence of a small population of unusually large vortices, which are likely associated with long-lived, large-scale coherent structures sustained by strong nonlinear interactions.

Overall, these results indicate that stronger self-interaction systematically suppresses small-scale rotational fluctuations, reduces the variability in vortex size, and enhances large-scale flow coherence. The simultaneous decrease in mean vorticity and the emergence of long-range velocity correlations point to a consistent physical mechanism: nonlinear self-interactions promote the self-organization of the flow into persistent, system-spanning circulation patterns, while efficiently damping turbulent motions at smaller scales. This interplay between interaction strength and flow organization appears to be a defining feature of three-dimensional quantum fluids.

\section{Conclusions}
\label{sec:conclusion}

In this study, we investigated the effects of self-interaction strength on the dynamics of soliton mergers in the ultra-light dark matter model, focusing on the formation and properties of turbulent vortex structures. Through numerical simulations of the Gross-Pitaevskii-Poisson (GPP) equations, we analyzed the impact of the self-interaction coupling parameter $g$ on vortex scales during soliton mergers. Our key findings, derived from velocity correlations, kinetic energy spectra, and vorticity distributions, reveal the significant role of self-interaction in shaping vortex formation and scaling, providing new insights into the small-scale structure of dark matter halos.

Our results confirm that vortex formation is a universal outcome of soliton mergers, independent of self-interaction strength. However, the spatial extent and intensity of these vortex structures are strongly modulated by $g$. Specifically, for repulsive self-interactions ($g > 0$), the vortex region expands as $g$ increases, leading to larger, more coherent vortical structures. Conversely, for attractive self-interactions ($g < 0$), the vortex region contracts, resulting in smaller, more compact vortices. This behavior is evident in the snapshots of vorticity and velocity fields, where the spatial distribution of vortices broadens with increasing $g$.

The velocity correlation function reveals a pronounced dependence on $g$. For negative or low $g$ values (e.g., $\widetilde{g} = -0.020$), the velocity field exhibits short-range correlations, with rapid decorrelation over small distances ($\widetilde{r} \approx 0.9$). As $g$ increases to positive values (e.g., $\widetilde{g} = +0.500$), the correlation length extends significantly ($\widetilde{r} \approx 3.0$), indicating enhanced spatial coherence and the emergence of large-scale, organized motion patterns. This suggests that stronger repulsive self-interactions promote collective flow organization, fostering persistent, domain-spanning circulation patterns.

The kinetic energy spectra further corroborate these findings. For negative $g$ (e.g., $\widetilde{g} = -0.020$), the spectra peak at high wavenumbers ($k_{\rm peak} \approx 3$) with high energy amplitude ($E(k_{\rm peak}) \approx 0.5$), reflecting dominant small-scale fluctuations. As $g$ increases to positive values (e.g., $\widetilde{g} = +0.500$), the spectral peak shifts to lower wavenumbers ($k_{\rm peak} \approx 2$) with reduced amplitude, indicating a suppression of small-scale energy and the dominance of large-scale coherent structures. This shift suggests an inverse energy cascade, where energy is transferred from small to large scales, driven by nonlinear self-interactions.

Vorticity distributions provide additional insight into the flow dynamics. The spatially averaged vorticity $\langle |\omega| \rangle$ decreases with increasing $g$, with a subtle non-monotonic bump at small positive $g$ due to transient increases in local shear. The probability distribution function (PDF) of vorticity magnitude remains Gaussian across all $g$, indicating statistical homogeneity despite the presence of coherent structures. The vortex size distribution narrows with increasing $g$, reflecting a well-defined characteristic vortex size and reduced small-scale variability, consistent with the growth of large-scale coherence.

The observed dependence of vortex size on self-interaction strength can be attributed to the interplay between the nonlinear self-interaction term and quantum pressure in the GPP equations. Stronger self-interactions enhance nonlinear effects, leading to more pronounced interference patterns and larger-scale vortex formation during mergers. This interplay drives the transition from small-scale, disordered fluctuations to large-scale, coherent vortical structures, highlighting the critical role of nonlinear dynamics in shaping turbulent flows.

The physical significance of studying turbulence in this context lies in its implications for understanding the structure and evolution of dark matter halos. Turbulent vortex structures serve as dynamical probes of the underlying physics, revealing how self-interactions influence the organization and energy distribution within the system. In the ultra-light dark matter model, these structures can act as gravitational seeds, potentially influencing the condensation of ordinary matter and forming observable ring-like structures\cite{Alvarez-Rios:2025ydz}. By quantifying the dependence of vortex scales on $g$, our study provides a framework for connecting microscopic interactions to macroscopic flow patterns, offering insights into the formation and stability of dark matter halos. Furthermore, the observed turbulence properties, such as extended correlation lengths and coherent structures for higher $g$, suggest potential observational signatures, such as distinct density or velocity patterns, that could be probed in astrophysical observations.

In ultra-light dark matter scenarios, wave interference substructures, including vortices, induce flux anomalies of 5-10\% in strongly lensed systems through gravitational lensing magnification effects and may produce time-varying deflections due to fast-moving vortex rings, thereby providing a unique probe to distinguish substructure power spectra from traditional subhalos. For  QCD axions or ultralight bosons, these vortices and broader interference patterns diminish signals in certain detection experiments (e.g., cavity haloscopes)\cite{Hui_2021}. Our findings highlight the critical role of nonlinear self-interactions in modeling soliton dynamics and their influence on the small-scale structure of dark matter halos, thus holding significant implications for future Earth-based, astrophysical, and cosmological surveys aimed at detecting novel boson particles across a wide range of scales. Future work will focus on exploring the long-term evolution of these vortex structures, their stability under varying initial conditions, and their implications for observational signatures in cosmological contexts.

\acknowledgments
We would especially like to thank Zhipan Li for his assistance in setting up the clusters. We would also like to thank Hongyi Zhang for constructive comments which helped to improve this paper. JC acknowledges the support from the Fundamental Research Funds for the Central Universities (SWU-KR22012, SWU-KT25031) and the Chongqing Natural Science Foundation General Project (2023NSCQ-MSX1929). This research was also facilitated by the computational resources in the School of Physical Science and Technology at Southwest University.

\bibliography{Ref}

\end{document}